# Ultra-fast relaxation of electrons in wide-gap dielectrics

H.-J. Fitting[a], , V.S. Kortov[b] and G. Petite[c]

[a]Physics Department, University of Rostock, Universitätsplatz 3, D-18051 Rostock, Germany
[b]Urals State Technical University, Mira street 19, RUS-620002 Ekaterinburg, Russia
[c]Laboratoire des Solides Irradies, Ecole Polytechnique, UMR CEA-DSM/CNRS, F-91128 Palaiseau Cedex, France

## Abstract

Low-energy electrons scattered in the conduction band of a dielectric solid should behave like Bloch electrons and will interact with perturbations of the atomic lattice, i.e. with phonons. Thus the phonon-based description of low-energy scattering within an energy band structure of a solid bears certain advantages against common free-electron scattering mechanisms. Moreover, the inelastic scattering is described by the dielectric energy loss function. With these collective scattering models we have performed the simulation of excited electron relaxation and attenuation in the insulator $SiO_2$. After excitation to a mean initial energy of several eV their energy relaxation occurs within a short time interval of 200 fs to full thermalization. There is a very rapid impact ionization cooling connected with cascading of electrons at the beginning during the first 10 fs, followed by much slower attenuation due to phonon losses in wide-gap dielectrics and insulators.

## 1. Introduction

Over more than three decades silicon dioxide stands for the most important dielectric and insulating material in microelectronics and optics. The favourable electronic and optical features of $SiO_2$ are based on the wide forbidden gap of 9 eV in the electronic energy band structure and the natural compatibility in the highly developed silicon technology.

Perfect thin dielectric and optical $SiO_2$ layers, often with a well-defined interface to the silicon substrate, are mostly investigated by optical and electron spectroscopy, i.e. photoluminescence and cathodoluminescence. The latter one allows a probing within smallest volumina in sub-micron dimension. For this reason the electron scattering and transport should be thoroughly understood. Therefore, the electronic excitation processes with rapid carrier excitation and relaxation should be investigated. The related fast conduction electron dynamics in optical breakdown processes under high-intensity laser excitation has been investigated in Ref. [1], and the photo- and exoelectron emission after radiation excitation and relaxation in Refs. [2] and [3]. The high-energy charge injection and the subsequent relaxation occur in ultra-short time scales and should be considered in the following.

# 2. Monte Carlo Simulation of impact ionization and phonon scattering

The full Monte Carlo (MC) program of low-energy electron scattering, especially in dielectric materials and insulators is mainly based on impact ionization and electron–phonon interaction and has been described in Refs. [4], [5] and [6].

## 2.1. Longitudinal optical (LO) phonons

The scattering frequency of electrons with energy $E$ in a parabolic band with LO phonons of energy $\hbar\omega_{LO}$ and Bose population NLO is given according to the Fröhlich theory, see, e.g. Ref. [7]:

$$f_{LO}^{\mp} = \frac{e^2}{4\pi\varepsilon_0 \hbar^2}\left(N_{LO} + \frac{1}{2} \mp \frac{1}{2}\right) \times \left(\frac{1}{\varepsilon_\infty} - \frac{1}{\varepsilon}\right)\sqrt{\frac{m^*}{2E}\hbar\omega_{LO}} \ln\left[\frac{1+\sqrt{1\pm\hbar\omega_{LO}/E}}{\mp 1+\sqrt{1\pm\hbar\omega_{LO}/E}}\right] \qquad (1)$$

where the upper signs stand for phonon absorption (−) and the lower ones for phonon emission (+) $m^*$ is the effective electron mass, $\varepsilon$ and $\varepsilon_\infty$ are the dielectric permittivities on the left- and right-hand side from the ionic polarization, i.e. the static and the optical dielectric constant, respectively. In $SiO_2$, we use $\varepsilon \cong 3.84$ and $\varepsilon_\infty = 2.25$ distributed to two LO modes $\hbar\omega_{LO}=150$ and 60 meV.

## 2.2. Acoustic (a.c.) phonons

After the optical runaway and heating-up to energies in the eV region the electrons feel a stronger acoustic (a.c.) coupling with the lattice. The energy loss or gain with some meV is much less than the energy of hot electrons. Therefore this a.c. scattering is considered as nearly elastic.

A strong increase of a.c. scattering is obtained at higher electron energies of some eV. In order to avoid irreal high a.c. scattering rates at high energies Bradford and Woolf [8] have introduced a Coulomb-like screening, leading to the following expressions:

$$f_{a.c.} = \frac{4\pi m_D(2N_{qBZ}+1)}{\rho \hbar \hbar \omega_{BZ}} E_{a.c.}^2 D(E) E \left(\frac{A}{E}\right)^2 \times \left[\ln\left(1+\frac{E}{A}\right) - \frac{E/A}{1+E/A}\right] \qquad (2)$$

with $D(E)$ as the density of states (DOS) and the effective electron mass $m_D$, $E_{a.c.}$ is the acoustic deformation potential, $A$ one screening parameter, $\rho$ the mass density, $E_{BZ}$ the electron energy at the Brillouin zone (BZ) boundary $k_{BZ}$, $N_{qBZ}$ and $\hbar\omega_{BZ}$ are acoustic phonon population and energy at the BZ boundary, respectively.

## 2.3. Impact ionization (i.i.)

At still higher electron energies $E>E_g$ exceeding the forbidden gap $E_g$ between valence and conduction band, direct valence band ionization, called impact ionization (i.i.), becomes possible.

Stobbe et al. [9] have calculated the transition rate between parabolic bands and found an appropriate formula:

$$f_{i.i.} = C \left[ \frac{E/E_{th}-1}{1+DE^2/E_{th}^2} \ln\left(\frac{E}{E_{th}}\right) \right]^a \quad for \quad E > E_{th} \quad (3)$$

This is an expression including the soft ionization threshold of Keldysh proportional to $(E-E_{th})^2$, as well as a hard threshold behaviour of the Bethe type: $\ln(E/E_{th})$.

In Fig. 1, the dielectric loss function $Im(-1/\varepsilon(\omega))$ due to excitation of LO phonons and to ionization of valence band electrons (i.i.) are presented in the upper part, further on the respective scattering rates of all mechanisms. Eqs. (1)–(3) are given in the lower part of Fig. 1. The detailed material scattering parameters can be found in Ref. [5].

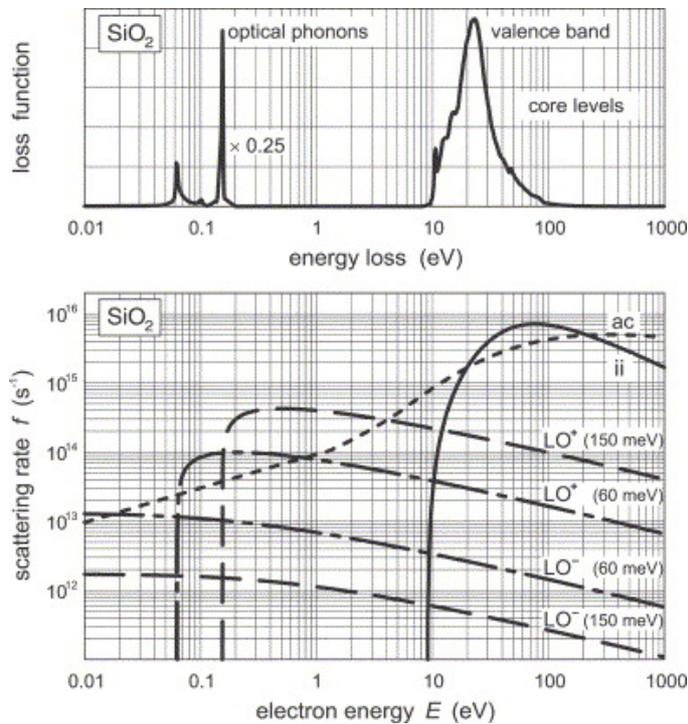

Fig. 1. Dielectric energy loss function (above) and corresponding scattering rates of electrons with longitudinal optical (LO) phonons of energies $\hbar\omega_1$=150 meV and $\hbar\omega_2$=60 meV; LO$^+$ phonon emission, LO$^-$ phonon annihilation; then with acoustic (a.c.) phonons, and due to impact ionization (i.i.) of valence band electrons in SiO$_2$ corresponding to the dielectric energy loss function above.

# 3. Fast electron relaxation in the conduction band

The development of new "pump and probe" techniques by means of ultra-short laser pulses allows to resolve the time-dependent relaxation of inner photoelectrons (PE) in photoluminescence (PL) and photoelectron emission spectroscopy [1] or electron beam-induced secondary electrons (SE) in cathodoluminescence (CL) spectroscopy [10]. According to the electron energy loss function in $SiO_2$ (Fig. 1) the excited inner SE possess initial energies of some tens of eV [4] and [5]. In the first step after creation the energy of SE is decaying rapidly due to very fast impact ionization and cascading under 10 eV after 10 fs. This process is demonstrated in Fig. 2 and Fig. 3.

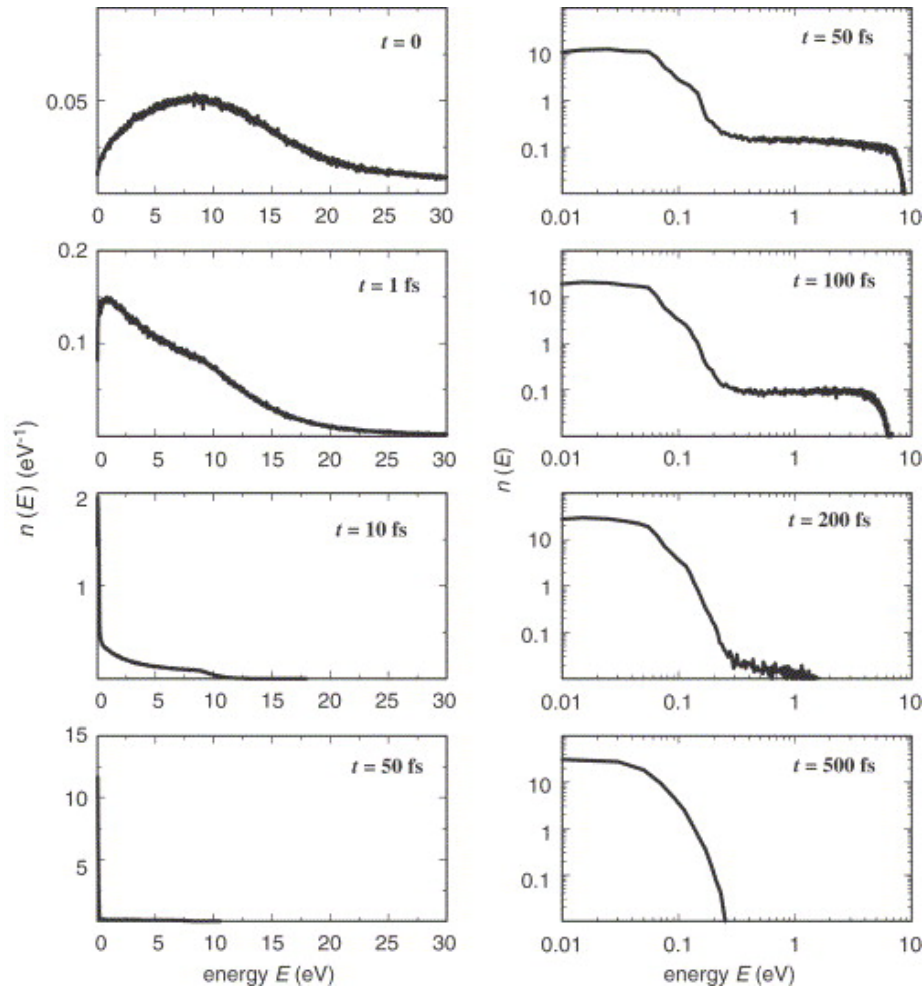

Fig. 2. Energy ralaxation of excited secondary electrons with time *t*. The rapid cooling at the beginning - *t*<10 fs - is due to impact ionization and cascading, the slower attenuation over 200 fs mainly due to the emission of LO phonons.

Afterwards the electrons are exposed to phonon interaction and are cooled down much slower till they are almost thermalized after about 200 fs. Their mean energy $\langle E \rangle$ in Fig. 3 (top) shows rapid cooling at the beginning due to impact ionization and cascading, then a much slower attenuation down to the thermalization at $\langle E \rangle \approx 40$ meV corresponding to the lattice temperature $T \approx 300$ K according to the Boltzmann statistics: $\langle E \rangle = 3/2\ kT$. This fairly instantaneous agreement is really surprizing. In the lower part of Fig. 3, the number *n(t)* of electrons related to their initial concentration $n_0$ is plotted versus time *t* and we clearly observe the three steps; cascading, attenuation, and thermalization again. Spontaneous thermalization below the lattice temperature, i.e. 40 meV, should lead to electron–hole recombination and the electron vanishes from the conduction band. So after about several ps we find the electrons as almost recombined or trapped. These processes occur over very short time scales in SiO$_2$ because of the high impact ionization rates with a great energy loss of $\Delta E > 9$ eV and the subsequent energy losses due to a relatively high LO phonon mode $\hbar\omega$LO=150 meV as described in Fig. 1 and demonstrated in Fig. 3.

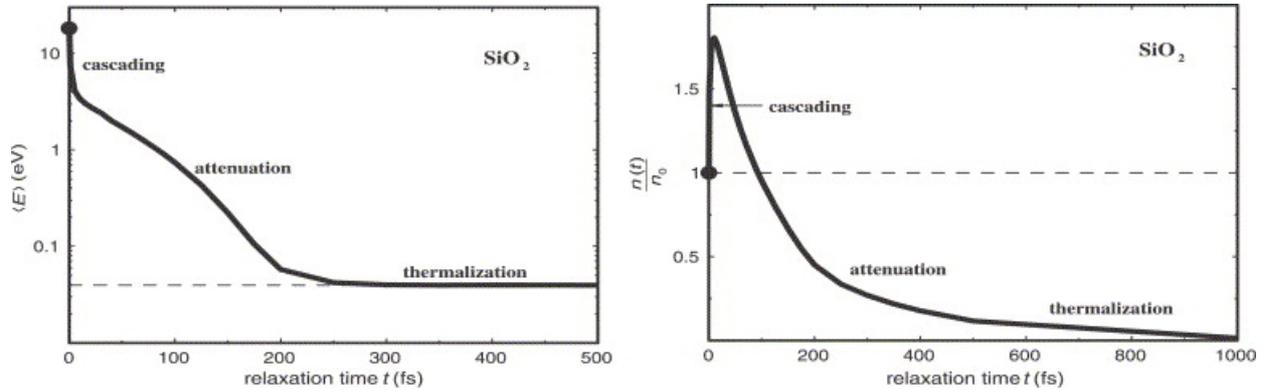

Fig. 3. Mean secondary electron (SE) energy <E> (left) and the SE rate $n/n_0$ (right) during the relaxation time $t$, with rapid cooling due to impact ionization and cascading at the beginning followed by slow attenuation due to phonon emission, final thermalization, and recombination.

## Acknowledgements

We are grateful to the support by the INTAS project 01-0458 "Fast electronic processes in dielectrics".